# Secure and Safety Mobile Network System for Visually Impaired People

Shyama kumari Arunachalam, Roopa V, Meena H B, Vijayalakshmi, T Malavika

MVJ College of engineering, Bangalore

*Abstract*- **The proposed system aims to be a techno-friend of visually impaired people to assist them in orientation and mobility both indoor and outdoor. Moving through an unknown environment becomes a real challenge for most of them, although they rely on their other senses. An age old mechanism used for assistance for the blind people is a white cane commonly known as walking cane a simple and purely mechanical device to detect the ground, uneven surfaces, holes and steps using simple Tactile-force feedback.**

I. INTRODUCTION

The Sesamonet system consists of a PDA/Smart phone processing unit, an earphone (Bluetooth or cable) and a special "RFID-aware" Bluetooth enabled walking cane. The RFIDs (Radio Frequency Identification) are inserted into the ground to form a virtual path to guide non/hypo-viewing people through different places (e.g. walking areas, working places, city centers, public buildings, etc.). The walking cane automatically reads the code written in the RFID and sends it to the cell phone or PDA of the visually impaired person. The PDA/Smart phone processes the received information by searching it in its internal database, associates it to a text message that is transformed in audio by a Text-to-Speech module and transmitted to the earphone.

The main encouraging factor for the application of technology for visually impaired people is the policy measures adopted by the western countries for social inclusiveness. Lot of development work in these countries is attributed to the above mentioned policy measures and grants invested for supporting the work. In the last year's international loss showcased many such devices. Some notable devices exhibited were PAN OPTICUS (sits between a digital satellite receiver and a TV to read the on-screen menus), MONOMOUSE (fitted with a diffuser in order to minimize glare when reading things like CD or DVD covers for low vision people), SONUS 1XT ( a device for voice readout of the scrolling text that most stations transmit to provide access of electronic resources to the visually impaired people by changing the formats of the websites and URLs. Yet another successful

Undertaken was RadioVirgilio/Sesamonet to design and implement a reliable system to assist visually impaired citizens' independent mobility in urban settings. The goal was achieved by integrating traditional assistive technologies with wireless and RFID technologies to realize an intelligent and easy to use navigation system. However the system was not very successful, since it failed to integrate with the state of art internet technology. Some new commercial devices appear on the market, like the Ultra Cane which uses a build-in sonar system and sends back vibrations through the handle according to the presence of obstacles. The ultra-cane enhanced the traditional white cane by giving information about the obstacles before direct contact. But it doesn't provide any new functionality to the traditional cane and the localization is still done by movement of the cane and it doesn't detect objects at head height. Most of the system developed so far focus on maintaining spatial orientation which is a major challenge for people with visual impairment. There is the need of systems in providing blind people with information on where they are hazards that might be in the way, and a description of what lies in their surroundings. The notion of "Spatial orientation" refers to the ability to establish awareness of space position relative to land marks in the surrounding environment. The goal of supporting functional independence to visually impaired people can be achieved by providing references and sorts of land marks to enhance awareness of the surroundings. Systems on similar lines have been reported and even patents are also filed in this area of development.

The literature survey reveals several following important trends:
Although the number of blind people is going up day by day, there are very few systems, publications or patents reported in recent years. In spite of lot of development work done in this area. None of the systems are successful and distributed worldwide. The reasons for this are two fold viz. techno-marketing and cost. In the first place for the blind people, it is hard to convenience them regarding the usefulness of the product due to their basic impairment. Secondly the costs of these devices are prohibitibly high and therefore they are out of the reach of the common man especially for the Asian countries. The above mentioned facts are the motivating factors behind taking up the proposed project.

## II. BLOCK DIAGRAM

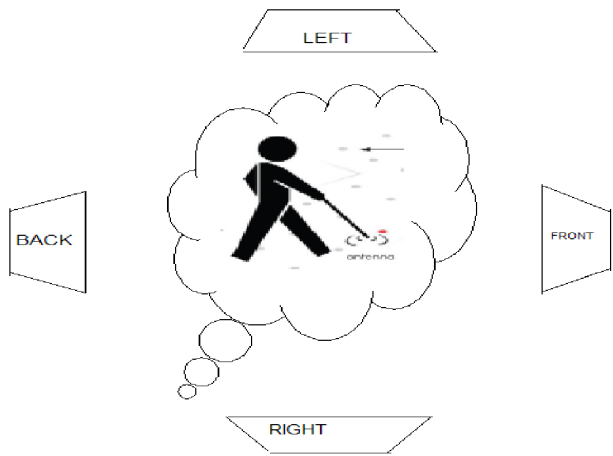

Figure 1: Block schematic of the proposed system Left, Right, Front and Back: Locations of arrays of RFID tags

## III. WORKING

In this project the user will have the cell phone which will be connected to the stick using the blue tooth protocol as shown in the block diagram above. The stick has the RFID and this is interfaced to the micro controller. A single row of RF tags will be positioned along a line, which follows a specific path, and embedded into a carpet. While the user walk along the path and read the RF tags with the walking cane. The micro controller reads the data and then sends the same to the cell phone. The application running in the cell phone will read the data and then it will play the respective message stored and this message also tell him what the person encounters in which ever direction the person is walking . While the costs of RFID tags are decreasing, this project kept costs at a minimum. Thus this way the user can be totally independent and can walk with confidence and with higher efficiency of his / her performance. In addition, information concerning the presence of shops or public transportation stops, but also giving information on the surrounding environment, warning for danger and avoiding obstacles, will be automatically provided. This will allow to "seeing through one's ears". If an emergency were to arise, the user can communicate with customer care unit with his exact position through a series of snaps to customer care unit, just pushing a specific button on the PDA/Smart phone. The person at customer care unit will see the snap and send the text inform to the cell phone , the cell phone will convert from text to audio , so that the person "see through one's ears".

## IV. DETAILED BLOCK DIAGRAM

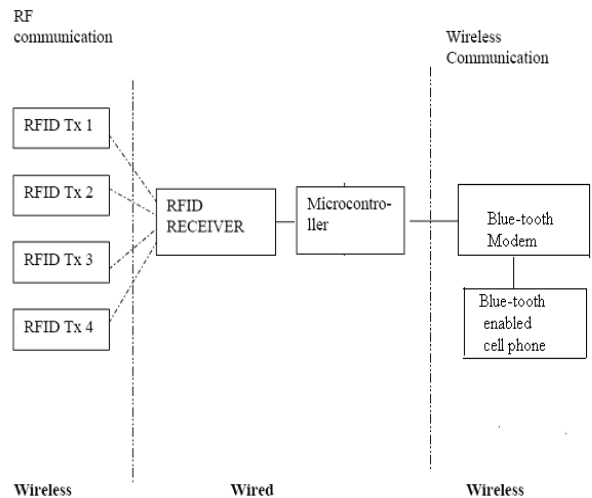

Figure 2: Detailed Block Diagram of the System

## V. FLOW CHART

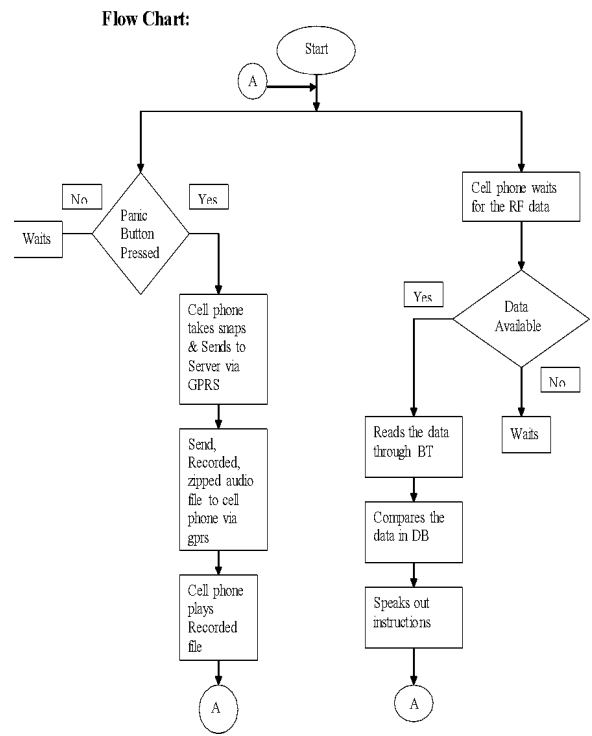

### DESCRIPTION

Initially the cell phone waits for the RF data i.e, RFID number, when the RF reader detects the RFID tag it sends that number to the cell phone through Bluetooth model. If not, waits for the data. In cell phone it compares with the database and speaks out the details that have been stored in the database.

If the person is not cleared or in case of emergency, he/she press the panic button if not, waits for the panic button to be pressed. Once the panic button is pressed the cell phone takes a series of snap shots and sends to the server via GPRS.

At the server, the server receives the images, views it and records the information about the picture and the compressed audio file is sent to the cell phone. In the cell phone the compressed audio is received and played.

## VI. Existing System

It is known that people who are blind/visually impaired find it difficult to move, especially in unknown places. Usually the only help they have is their walking stick (white cane), a guide dog and sometimes special warning sounds or road signals at specific positions.

## VII. Innovative Features Of The Project

- The proposed system is innovative because so far no commercial organization or research group has come forward to integrate the benefits of the Bluetooth protocols and the RFID tags to help the visually impaired people.
- Project is taking the benefit of the fact that the Passive or active RFID tags are becoming ever less expensive and easily available.
- Most of the systems reported so far are based on PCs. Proposed system is based on microcontroller which not only optimizes the space but also power and brings down the cost drastically.

## VIII. Proposed System

The SESAMONET system uses RFID technology for user localization and tracking. SESAMONET uses a grid of RFID tags which are burrowed in the ground. An RFID reader is attached to a cane to obtain the tag ID as the cane moves over the tag. This information is sent to a PDA where software looks up the navigation data for the tag ID. The navigation data is converted to speech using text-to-speech synthesis. The proposed system is innovative because so far no commercial organization or research group has come forward to integrate the benefits of the Bluetooth and the RFID tags to help the visually impaired people. Project is taking the benefit of the fact that the Passive or active RFID tags are becoming ever less expensive and easily available. Most of the systems reported so far are based on PCs. Proposed system is based on microcontroller which not only optimizes the space but also power and brings down the cost drastically. In case of uncomfortable situation, the blind person takes a series of snap shot of the surrounding and sends it to customer care unit. The transcribers will convert the image information to text and send to the cell phone of a blind person. The application will convert the text to audio and play for the person.

## IX. Hardware Requirement

The hardware requirements for this project are

RFID Reader

Microcontroller

Bluetooth

Power Supply

RFID tags

## X. Software Requirement

The software requirements for this project are

Operating System: Windows versions

Java Communication API

Hyper terminal

Android OS 2.2

## XI. Advantages

- Cost effective and time efficient.
- Reduces the risks.
- Easy to construct and install.
- Consumes less energy and is more efficient.
- Increases the overall efficiency of the person.
- Works at high speeds etc.

## XII. Applications

- Used when working alone
- Used when needs to know about the surroundings.
- Used within campus or office premises etc.

XIII. Conclusion

We can get the benefit of the RFID tags which are becoming ever less expensive and easily available. This system is based on microcontroller which do not only optimizes the space but also power and brings down the cost drastically.

XIV. REFRENCES